\documentclass[a4paper,12pt]{ETHpaper}

	\usepackage{graphicx}
	\usepackage[dvipsnames]{xcolor}
	\usepackage{amsthm}
	\usepackage{booktabs}
	\usepackage{multirow}
	\usepackage{arydshln}
	\usepackage{tikz}
	\usetikzlibrary{arrows.meta}
	\usetikzlibrary{calc}
	\usepackage{pgfplots}
    \pgfplotsset{compat=newest}
	\usepgfplotslibrary{fillbetween}

	\usepackage[sort]{natbib}
	\usepackage[toc,nonumberlist]{glossaries}

	\newcommand{\localtimevar}{t}
	\newcommand*\kin{\ensuremath{k^{\text{in}}}}

	\newcommand*\crate{\ensuremath{\tilde{c}}}
\newcommand{\effectslegend}{
	\newcommand{\legendnode}{
		\tikz[baseline={([yshift=-0.7ex]current bounding box.center)}]{
			\node[rectangle, fill=c1, inner sep = 0pt, minimum size=1em] at (0, 0){};
			\node[rectangle, fill=c2, inner sep = 0pt, minimum width=0.9em, minimum height=0.2em] at (0, 0){};
		}
	}

	\definecolor{c1}{HTML}{fde3e1}
	\definecolor{c2}{HTML}{faa6a0}
	\legendnode{}
	\gls{jpr}
	\ \ \definecolor{c1}{HTML}{f6e9cc}
	\definecolor{c2}{HTML}{e3ba5d}
	\legendnode{}
	\gls{jpra}
	\ \ \definecolor{c1}{HTML}{e9eecc}
	\definecolor{c2}{HTML}{b1c247}
	\legendnode{}
	\gls{jprc}
	\ \ \definecolor{c1}{HTML}{ccf1d7}
	\definecolor{c2}{HTML}{4fcf75}
	\legendnode{}
	\gls{jpre}
	\ \ \definecolor{c1}{HTML}{ccf2eb}
	\definecolor{c2}{HTML}{6bdbc7}
	\legendnode{}
	\gls{jrmp}
	\ \ \definecolor{c1}{HTML}{ccf1f9}
	\definecolor{c2}{HTML}{38c8e9}
	\legendnode{}
	\gls{jhep}
	\ \ \definecolor{c1}{HTML}{dfebff}
	\definecolor{c2}{HTML}{649eff}
	\legendnode{}
	\gls{jpr_ins}
	\ \ \definecolor{c1}{HTML}{f7e2fe}
	\definecolor{c2}{HTML}{e394fc}
	\legendnode{}
	\gls{jpl}
	\ \ \definecolor{c1}{HTML}{ffdff3}
	\definecolor{c2}{HTML}{ffa6dd}
	\legendnode{}
	\gls{jnp}
}

\newacronym{j1213103}{JHEP}{The Journal of High Energy Physics} \newacronym{j1214516}{PR-HEP}{High Energy Physics in Physical Review Journals} \newacronym{j1214521}{Phys.\,Lett.}{Physics Letters} \newacronym{j1214548}{Nuc.\,Phys.}{Nuclear Physics} \newacronym{jhep}{JHEP}{The Journal of High Energy Physics} \newacronym{jpr_ins}{PR-HEP}{High Energy Physics in Physical Review Journals} \newacronym{jpl}{Phys.\,Lett.}{Physics Letters} \newacronym{jnp}{Nuc.\,Phys.}{Nuclear Physics} 
\newacronym{aps}{APS}{American Physical Society}
\newacronym{j0}{PR}{Physical Review} \newacronym{j1}{PRA}{Physical Review A} \newacronym{j3}{PRB}{Physical Review B} \newacronym{j4}{PRC}{Physical Review C} \newacronym{j5}{PRD}{Physical Review D} \newacronym{j6}{PRE}{Physical Review E} \newacronym{j7}{PRI}{Physical Review I} \newacronym{j8}{PRL}{Physical Review L} \newacronym{j9}{PRSTAB}{Physical Review STAB}  \newacronym{j10}{PRSTREP}{Physical Review STREP}  \newacronym{j12}{RMP}{Reviews of Modern Physics}  \newacronym{jpr}{PR}{Physical Review} \newacronym{jpra}{PRA}{Physical Review A} \newacronym{jprb}{PRB}{Physical Review B} \newacronym{jprc}{PRC}{Physical Review C} \newacronym{jprd}{PRD}{Physical Review D} \newacronym{jpre}{PRE}{Physical Review E} \newacronym{jpri}{PRI}{Physical Review I} \newacronym{jprl}{PRL}{Physical Review L} \newacronym{jprstab}{PRSTAB}{Physical Review STAB}  \newacronym{jprstrep}{PRSTREP}{Physical Review STREP}  \newacronym{jrmp}{RMP}{Reviews of Modern Physics}

\begin{document}

\title{Citations Driven by Social Connections?\\ A Multi-Layer Representation of Coauthorship Networks}  
\author{Christian Zingg\(^*\) \qquad Vahan Nanumyan\(^\dagger\) \qquad Frank Schweitzer\(^1\)}
\authoralternative{Ch.~Zingg, V.~Nanumyan, F.~Schweitzer}
\address{Chair of Systems Design, ETH Zurich, Switzerland \\[2mm]
\(^*\)czingg@ethz.ch\\
\(^\dagger\)vahan@nanumyan.com\\
\(^1\)fschweitzer@ethz.ch
}
\www{\url{http://www.sg.ethz.ch}}
\reference{(Submitted for publication)}\maketitle

\begin{abstract}
	To what extent is the citation rate of new papers influenced by the past social relations of their authors?
	To answer this question, we present a data-driven analysis of nine different physics journals.
	Our analysis is based on a two-layer network representation constructed from two large-scale data sets, INSPIREHEP and APS.
	The social layer contains authors as nodes and coauthorship relations as links.
	This allows us to quantify the social relations of each author, prior to the publication of a new paper.
	The publication layer contains papers as nodes and citations between papers as links.
	This layer allows us to quantify scientific attention as measured by the change of the citation rate over time.  We particularly study how this change depends on the social relations of their authors, prior to publication.
	We find that on average the maximum value of the citation rate is reached sooner for authors who either published more papers, or who had more coauthors in previous papers.
	We also find that for these authors the decay in the citation rate is faster, meaning that their papers are forgotten sooner.

	\vspace{1em}
\end{abstract}

\glsunset{jpr}
\glsunset{jpra}
\glsunset{jprc}
\glsunset{jpre}
\glsunset{jrmp}

\glsunset{jhep}
\glsunset{jpr_ins}
\glsunset{jpl}
\glsunset{jnp}

\section{Introduction}\label{sec:attention:introduction}

The availability of large-scale data sets about journals and scientific publications therein, their authors, institutions, cited references and citations obtained in other papers has boosted scientometric research in the past years.
They allow to address new research questions that go beyond the calculation of mere bibliographic indicators.
These regard particularly the role of \emph{social influences} on the success of papers, for example coauthorship relations \citep{Sarigol2014} or the relations between authors and handling editors \citep{Sarigol2017}.
Such investigations have contributed to a new scientific discipline, \emph{the science of success} \citep{Sinatra2018, Jadidi2018}.

But such data also allows to redo traditional scientometric analyses on a much larger scale.
In \citep{Parolo2015}, the dynamics of the citation \emph{rate}, i.e. the change in the number of citations during a fixed time interval, is analysed.
The authors find that the change of the average citation rate follows two characteristic phases, first a growth phase and then a decay phase.
Interestingly, the duration of the first and the speed of the second phase have changed over the years.
This allows conclusions of how the \emph{collective attention} of scientists towards a given paper has evolved between early and recent times.

In general, the dynamics of citations are extensively studied in the bibliometric literature.
For example, the relation between the current number of citations and the citation rate was studied in \citep{Jeong2003}.
Or, citations were found to occur in bursts, with large bursts within few years after publication \citep{Eom2011}.
Concerning the scientific field of a paper, citations from papers in the same field tend to be obtained earlier than citations from papers in other fields \citep{Rinia2001}.
Different from such studies, citation rates were also used to classify papers \citep{Avramescu1979, Li2014a}.
Such classes often identify papers that receive citations earlier or later than the majority of papers \citep{Costas2010, Ciotti2016, Colavizza2016}.
Papers in the second class, i.e. which receive their citations only a long time after publication, are often called ``sleeping beauty'' or ``delayed'' \citep{van2004sleeping,burrell2005sleeping}.
Their citation rate and how it differs from other papers was studied extensively in \citep{Lachance2014}.
This class has been thoroughly studied also outside paper classification settings.
It was found that ``sleeping beauties'' are extremely rare, and only 0.04\% of papers published in 1988 were identified as such \citep{van2004sleeping}.
They were also found to occur especially often in multi-disciplinary data sets \citep{Ke2015}.

The recent progress in the study of scientometric systems very much relies on representing them as networks.
A first example are \emph{citation networks} representing papers as nodes and citations as their (directed) links.
Such networks can be seen as a knowledge map of science \citep{Leydesdorff2013}.
They can be also used to predict scientific success \citep{Mazloumian2012}.
A second example are \emph{coauthorship networks} representing scientists as nodes and their coauthorships as links.
While sociological studies \citep{Cetina2009} just report that communication between coauthors can be very intricate,
also formal models of how such collaborations form on the structural level have been developed \citep{Guimera2005, Tomasello2017}.
To study collaboration patterns in a university faculty \citep{Claudel2017}, such coauthorship networks have been combined with a network encoding the physical distance between the faculty members. It was also analysed how communities detected on a coauthorship network  overlap with different research topics \citep{Battiston2016}.

These investigations have the drawback that they study citation networks and coauthorship networks separately from each other.
As already emphasized \citep{Clauset2017,Schweitzer2014}, this becomes a problem if one wants to study social influence on citation dynamics.
For example, based on a data set of \emph{Physical Review} it was shown that scientists cite former coauthors more often \citep{Martin2013}.
Therefore, a better approach is to combine both the citation and the coauthorship network in a \emph{multilayer network}.
Links between the citation and the coauthorship layer express the authorship of papers.
Using such a representation, a method to detect citation cartels was proposed \citep{Fister2016}.
Further, the rate of citations dependent on the authors' total number of citations was studied \citep{Petersen2014}.
However, it was not investigated yet how the \emph{position} of authors in the coauthorship network influences \emph{when} their papers are cited.
In this paper we study exactly this question.

Our analysis extends recent studies that focus on the success of papers as measured by their total number of citations.
In \citep{Sarigol2014}, this success was related to the position of the authors in a coauthorship network.
It was shown that authors of successful papers are considerably more central (as quantified by \emph{various} centrality measures) in the coauthorship network.
We extend this by an analysis of the dynamics of the citation rate over time, i.e. \emph{when} their papers are cited.
To parametrise the citation dynamics, we resort to the  mentioned phases identified in \citep{Parolo2015}.
We extend this work by relating these phases to the social relations of the authors.

Our paper is structured as follows.
In Sect.~\ref{sec:attention:cit-rate} we explain how citation dynamics can be measured by means  of \emph{citation histories}, which represent the collective attention.
In Sect.~\ref{sec:bibl-datab} we describe the data sets used for our analysis.
In Sect.~\ref{sec:attention:multilayer-representation} we introduce the multilayer network to combine social information about authors with citation data.
We then turn to our research question and study in Sects.~\ref{sec:attention:peak-time}, \ref{sec:attention:decay} how the social relations of authors in the coauthorship network influence the collective attention.
Lastly, in Sect.~\ref{sec:attention:conclusion} we conclude our findings.

\section{Methods and Data}
\label{sec:data}

\subsection{Dynamics of Citation Rates}
\label{sec:attention:cit-rate}

\paragraph{Measuring attention. }

Citations are often used as a measure of \emph{success} of a paper, accumulated over time.
They have the advantage that they are objective in the sense that they are protocolled in the reference lists of citing papers.
But the sheer number of citations does not utilize the \emph{temporal information}, i.e. how many of these citations arrive at a given time.
This is captured in the citation \emph{rate}, which better estimates the \emph{attention} a paper receives at a given time (interval).
Individual attention, i.e. \emph{who} cites a given paper at a given time, is not of interest for our study.
We focus on \emph{collective} attention, i.e. aggregate over all authors who cite this paper during a given time interval.
Obviously, the citation rate is only a proxy of this collective attention.
One could additionally consider other attention measures like the \texttt{altmetric} score.
But such information is only available for very recent publications and further strongly biased against the use of social media.
Therefore, we decide to restrict our study to only using the citation rate as a proxy for collective attention.
At least, most papers are still cited because they have caught in some way the attention of the authors of the citing papers.
Furthermore, citation counts were found to be a good approximation of scientific impact as perceived by scientists from the same field as the paper \citep{Radicchi2017a}.

\paragraph{Citation histories. }
We measure the collective attention of a paper by the number of citations it receives over a particular time interval, i.e. its citation rate.
More precisely, for paper $i$ published at time $\delta_i$, the citation rate at \(t = \delta - \delta_i\) time units after publication is

\begin{equation}
    c_i(t) = \frac{\kin_i(\delta + \Delta t) - \kin_i(\delta)}{\Delta t}
    \label{eq:attention:cit-rate-i}
\end{equation}

where \(\kin_i(\delta)\) denotes the total number of ``incoming'' citations the paper has received at time \(\delta\).
The dynamics of the citation rate \(c_i(t)\) is also called \emph{citation history} of paper \(i\)~\citep{Parolo2015}.
To compare  citation histories across papers we further normalise them by their respective maximum value \(c_i^{\max} = \max_t\{c_i(t)\}\):

\begin{equation}
    \crate_i(t) = c_i(t) / c_i^{\max}.
    \label{eq:attention:norm-cit-rate-i}
\end{equation}

\paragraph{Two phases in citation histories. }
\citet{Parolo2015} find two characteristic phases in the dynamics of normalised citation histories \(\crate_i(t)\) of a paper $i$.
In the first phase, which lasts for 2--7 years, it grows and eventually reaches a peak at a time $t^\text{peak}_i$.
After the peak there is the second phase, in which the citation rate decays over time.
For the majority of papers this decay was found to be well described by an exponential function:

\begin{equation}\label{eq:attention:exp-decay}
    \crate_i(t) \propto \exp(-t/\tau_i),
\end{equation}

\begin{figure}[htbp] \centering
	\resizebox{0.6\textwidth}{!}{\definecolor{customG}{HTML}{218380}  \definecolor{customB}{HTML}{2E5EAA}  

\begin{tikzpicture}
    \tikzstyle{axis}=[]
    \tikzstyle{curve}=[thick]

    \coordinate (AxisOrigin) at (0, 0);
    \coordinate (AxisXMax) at (4, 0);
    \coordinate (AxisYMax) at (0, 2.5);
    \coordinate (AxisYPeak) at (0, 1.7);
    \draw[-latex, axis, name path=xAxis] (AxisOrigin) -- ($(AxisXMax)+(0.05, 0)$);
    \draw[-latex, axis, name path=yAxis] (AxisOrigin) -- ($(AxisYMax)+(0, 0.05)$);

    \node[below] (LabelOrigin) at (AxisOrigin) {\scriptsize $\color{black} 0$};
    \node[below] at (AxisXMax) {\scriptsize $\color{black} \localtimevar$};
    \node[left] at (AxisYMax) {\scriptsize $\color{black} \tilde{c}_i(\localtimevar)$};

    \draw ($(AxisYPeak) + (0.05, 0)$) -- ($(AxisYPeak) + (-0.05, 0)$);
    \node[left] at (AxisYPeak) {\scriptsize $\color{black} 1$};

    \coordinate (AxisXPeak) at (1, 0);
    \draw[name path=LinePhase1, curve, color=customG] (AxisOrigin) -- ($(AxisXPeak)+(AxisYPeak)$);
    \path[name path=xAxisPhase1] (AxisOrigin) -- (AxisXPeak);
    \node[anchor=center, xshift=5] at (AxisXPeak |- LabelOrigin) {\scriptsize $\color{customG} t_i^\text{peak}$};

    \draw[name path=LinePhase2, curve, color=customB] ($(AxisXPeak)+(AxisYPeak)$)
        to [out=290, in=178]
            node[midway, above right] {\scriptsize $\color{customB} e^{-\localtimevar/\tau_i}$}
            ($(AxisXMax)+(AxisOrigin)!0.05!(AxisYPeak)$);
    \path[name path=xAxisPhase2] (AxisXPeak) -- (AxisXMax);

    \coordinate (Phase1LabelLeft) at ($(AxisOrigin) + (AxisYPeak) + (0, 0.4)$);
    \coordinate (Phase1LabelRight) at (AxisXPeak |- Phase1LabelLeft);
    \coordinate (Phase2LabelLeft) at (Phase1LabelRight);
    \coordinate (Phase2LabelRight) at (AxisXMax |- Phase1LabelLeft);
    \draw[Latex-Latex, color=customG] (Phase1LabelLeft) -- node[midway, above] {\tiny \textcolor{customG}{Phase 1}} (Phase1LabelRight);
    \draw[Latex-Latex, color=customB] (Phase2LabelLeft) -- node[midway, above] {\tiny \textcolor{customB}{Phase 2}} (Phase2LabelRight);
    \draw[axis, dashed, color=lightgray] ($(Phase1LabelRight) + (0, 0.15)$) -- ($(AxisXPeak) + (0, -0.1)$);
    \draw[axis, dashed, color=lightgray] (AxisOrigin |- AxisYPeak) -- (AxisXMax |- AxisYPeak);

    \tikzfillbetween[of=xAxisPhase1 and LinePhase1]{customG, opacity=0.1};  \tikzfillbetween[of=xAxisPhase2 and LinePhase2]{customB, opacity=0.1};  \end{tikzpicture}}
    \caption{Illustration of the two characteristic phases in normalised citation histories $\crate_i(t)$ of most papers.
		}\label{fig:attention:history-two-phases}
\end{figure}
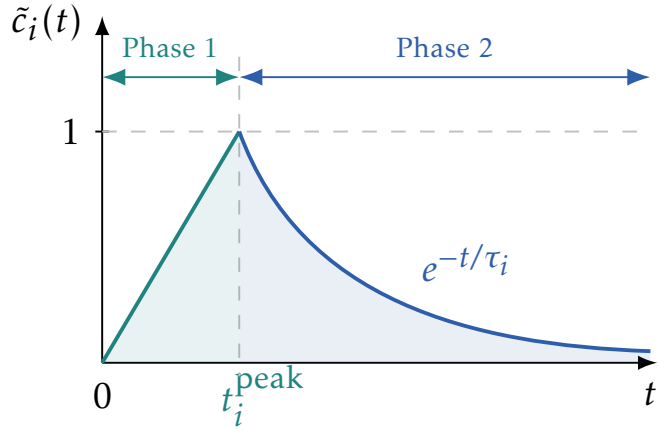

The parameter $\tau_i$ is called ``lifetime'', and it determines the speed of the decay.
The \emph{larger} $\tau_i$ is, the \emph{faster} is the decay.
Figure~\ref{fig:attention:history-two-phases} illustrates the two phases of $\crate_i(t)$.

\subsection{Bibliographic databases}
\label{sec:bibl-datab}

As we argued in Sect.~\ref{sec:data}, citations are particularly suitable to quantify the collective attention by scientists from the same field as a given paper.
Therefore, in our analysis we study different journals separately, because each describes a topic-related community of authors and their papers.
To obtain the data for our study we resort to large bibliographic databases which index papers across journals.
They collect information such as a paper's title, the list of authors, the date of publication, and also the list of references that a paper cites.
We extracted this set of information for $9$ journals from two such databases in the same way as in \citep{Nanumyan2020} and as explained below.

\paragraph{APS database.}
It indexes papers published in journals by the American Physical Society (APS).
Access to the database can be requested for research purposes at \url{https://journals.aps.org/datasets}.
We extracted the journals Physical Review (\gls{jpr}), Physical Review A (\gls{jpra}), Physical Review C (\gls{jprc}), Physical Review E (\gls{jpre}), and Reviews of Modern Physics \gls{jrmp} to cover a wide range of physics sub-fields.

The APS database has the known issue of name disambiguation, because it indexes authors by their name and not by a unique identifier.
This means that different authors with the same name are indexed as one author.
Such a ``multi-author'' then owns all papers and coauthorships that were actually accumulated by multiple authors.
In contrast, one author whose name can be spelled in different ways, may be indexed as different authors in the database.
The consequence for our study is that such undisambiguated authors bias measures involving (co-)authorships.
This problem was already discussed in the scientific literature, and a disambiguation algorithm specifically for authors in the APS~database was proposed \citep{Sinatra2016}.
We applied this algorithm to the APS~database to lower the bias from undisambiguated authors.

\paragraph{INSPIREHEP database.}
The second database, called INSPIREHEP, indexes papers relevant for the field of high-energy physics.
This database can be downloaded at \url{http://inspirehep.net/dumps/inspire-dump.html}.
In this database authors are disambiguated, because each author is indexed by a unique identifier.
We extracted the journals Journal of High Energy Physics (\gls{jhep}), Physics Letters (\gls{jpl}), Nuclear Physics (\gls{jnp}), and high energy physics literature in Physical Review journals (\gls{jpr_ins}) from this database.
They were the four largest journals in terms of number of citations from papers in the same journal, i.e. the citations which we will use to compute citation rates in the later Sections.

In INSPIREHEP some indexed papers have exceptionally large lists of authors, which sometimes even exceed $1000$ authors.
Such large-scale coauthorships were termed \emph{hyperauthorships} in \citep{Cronin2001}.
The concerns were raised, that it is unclear which authors actually made substantial contributions to such papers \citep{Cronin2001}, and that the coauthorship network is no accurate representation of the social network of authors \citep{Newman2004}.
Indeed, every author in such a hyperauthorship gets possibly thousands of collaborators from just a single paper, despite likely not having collaborated with all of them personally.
This introduces a bias for measures involving coauthorships, and thus for our study.
It was found that hyperauthorships usually occur in papers from large experiments \citep{Newman2001}, such as the ATLAS experiment at CERN.
To avoid this bias we remove experimental papers from the database.
To identify experimental papers we used meta-tags that INSPIREHEP provides, so-called XML-tags.
These are essentially labels for papers, which provide additional information such as arXiv identifiers, author affiliations, or sometimes estimates whether a paper is experimental or theoretical.
We removed all papers from the database which are explicitly tagged as \emph{experimental}.
But because this tag might be unavailable for a paper, we further removed all papers that are not explicitly tagged as \emph{theoretical work} or \emph{work in general physics}.

\begin{table}[htbp]
    \centering
	\caption{Overview of the extracted journals from the APS database and the INSPIREHEP database (IH).
		$|V^{p}|$ is the number of papers, $|V^{a}|$ is the number of authors, $|E^{pc}|$ is the number of citations between papers, and $|E^{a}|$ is the number of authorships.
	}
    \begin{tabular}{ccrrrr}
        \toprule
        Database & Journal & \(|V^{p}|\) & \(|V^{a}|\) & \(|E^{pc}|\) & \(|E^{a}|\) \\
        \midrule
        \multirow{5}{*}{APS} & \gls{jpr}   &  46728  &  24307  &  253312  & 87386 \\
        & \gls{jpra}  &  69147  &  41428  &  416639  & 144806 \\
        & \gls{jprc}  &  36039  &  22672  &  253948  & 108844 \\
        & \gls{jpre}  &  49118  &  36382  &  182701  & 95796 \\
        & \gls{jrmp}  &  3006  &  3788  &  5282  & 5044 \\
        \midrule
        \multirow{4}{*}{IH} & \gls{jhep}  &  15739  &  7994  &  191990  & 39056 \\
        & \gls{jpr_ins}  &  44829  &  33908  &  213625  & 115237 \\
        & \gls{jpl}  &  22786  &  18078  &  56332  & 53089 \\
        & \gls{jnp}  &  24014  &  18733  &  125252  & 60018 \\
        \bottomrule
    \end{tabular}\label{tab:dm:sum}
\end{table}

To summarize, Table~\ref{tab:dm:sum} provides summary statistics of the mentioned $9$ journals.
It further also shows how large these journals actually are.
For example, there is only $1$ journal, \gls{jrmp}, which contains less than $10000$ authors, or there are more than $400000$ citations between papers in \gls{jpra}.

\section{Social Influence on Citation Rate}\label{sec:attention:soc-inf}

\subsection{Multilayer Network Representation}\label{sec:attention:multilayer-representation}

\paragraph{Combining information about papers and authors. }

Our aim is to combine the information about collective attention, as proxied by the citation rate, with information about the social relations between authors.
For the latter, we specifically focus on \emph{coauthorship}, because this is the most objective and best documented relation.
Again, this is a proxy because it neglects other forms of social relationships, such as friendship, personal encounters, e.g. during conferences, electronic communication, or relations in social media.
But we do not have this type of information available for all authors over long times.
Therefore we restrict our analysis to the coauthorship network that can be constructed from available data, as described below.

To relate information
about authorship and about papers in a tractable manner, multilayer networks come into play, because they allow us to represent such separate information in different layers.
The nodes on the first layer correspond to papers and the (directed) links to their citations.
Different from this, the nodes in the second layer correspond to the authors and the links to their coauthorships, i.e., there is a link between two authors if they wrote at least one paper together.
Then, there are links which connect nodes on the first layer with nodes on the second layer.
These links correspond to the authorship relations, i.e., for every author there is exactly one such link to each of her papers.
We construct such a two-layer network for each of the $9$ journals in our data set to represent the information about citations between papers as well as about the authorships.

\begin{figure}[htbp] \centering
    \includegraphics[width=0.85\textwidth]{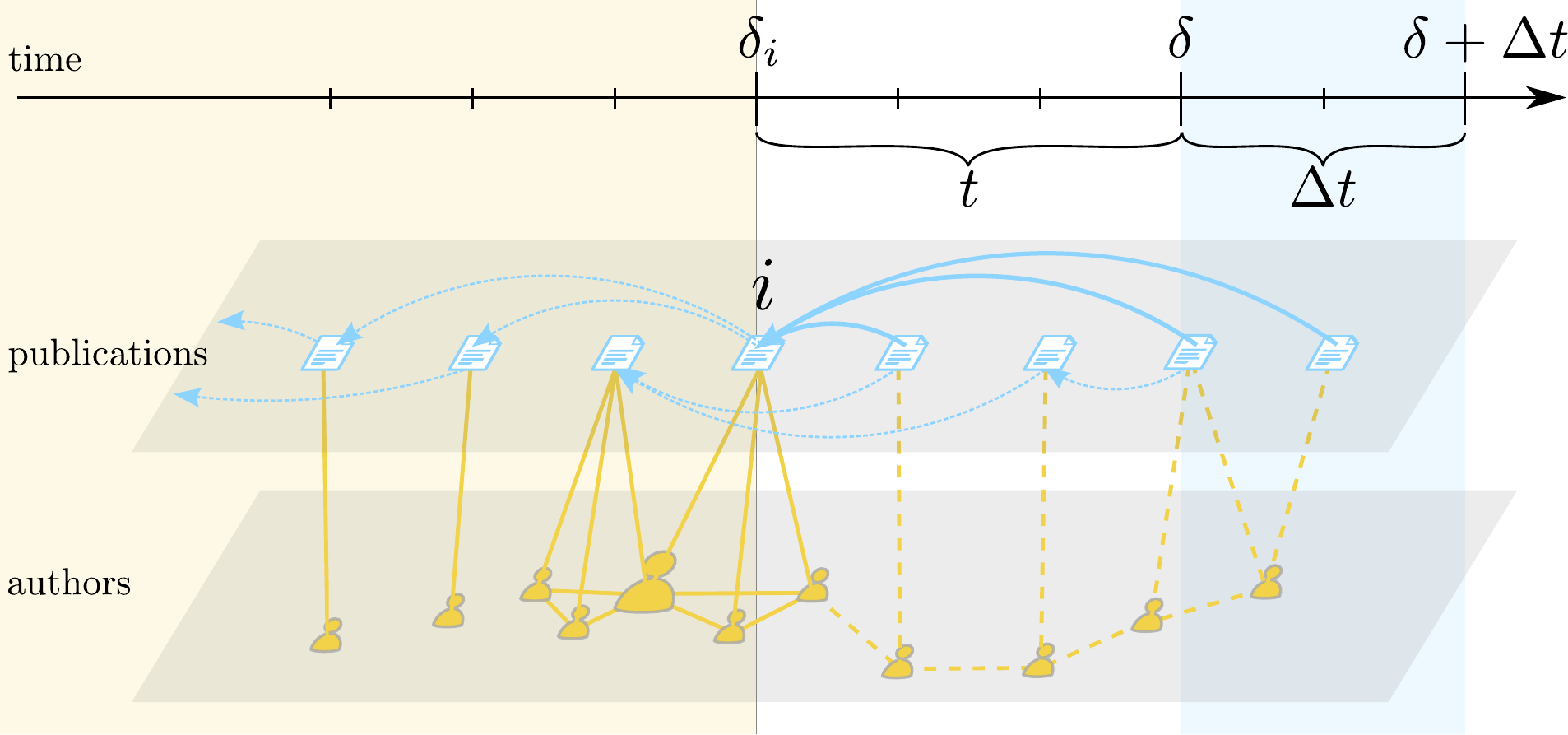}
    \caption{
      Multilayer network illustrating the coupling between the coauthorship network and the citation network.
Links between the two layers represent the relation between authors and papers.
      The timeline on top indicates that links within the citation layer are directed and point to papers already existing at the time when a paper is published. }\label{fig:attention:timeline}
\end{figure}

To summarize the above, Figure~\ref{fig:attention:timeline} illustrates the two layers of citation and coauthorship networks and their coupling.
It further displays the temporal dimension:
The multilayer network evolves over time because new papers are published, and hence new coauthors appear.
As the timeline indicates, paper $i$ is published at time $\delta_i$ and then accumulates citations in the future, at times $\delta>\delta_i$.
The publication layer allows us to define the degree of a \emph{paper} $i$ as the number $k_{i}^{\mathrm{in}}(\delta)$ of papers that cite $i$ until time $\delta$, see Eq.~\eqref{eq:attention:cit-rate-i} and Figure \ref{fig:attention:timeline}.
Specifically, it is the \emph{in-degree} because the publication network is directed.
The question is now how the citation rate of this paper evolves over time, conditional on the social information about its authors at time $\delta_i$, which is the publication time of paper $i$.
In other words, we analyse the impact of information from \emph{before} this publication.

\paragraph{Quantifying authors' social relations.}
The coauthorship layer allows us to define the \emph{degree} of an \emph{author} $n$ as the total number of
distinct coauthors $k_{n}(\delta_i)$ that the author had  \emph{before} time $\delta_i$.
Degree is the simplest centrality measure for networks and reflects the \emph{local} information about the embedding of an author in the social network.
We use it here because it was shown recently \citep{Nanumyan2020} that this measure is a particularly good predictor for the future citation rate.

We characterize a paper $i$ at time $\delta_i$ as the number of distinct coauthors of its authors \emph{before} time $\delta_i$.
This means that we sum up over the authors' individual degrees $k_{r}$

\begin{align}
  \label{eq:1}
  s^{NC}_i(\delta_i) = \sum_{r} k_{r}(\delta_i) - C(\delta_i)
\end{align}
and subtract a correction term $C(\delta_i)$ to not count those co-authors multiple times who collaborated with more than one author in the past.
The derivation of this correction term can be found in Appendix~\ref{sec:correction-terms}.
The index $NC$ refers to number of coauthors.
Furthermore, the paper $i$ published at time $\delta_i$ is \emph{not} counted in  $s^{NC}_i(\delta_i)$.

We also make use of the coupling between the two layers for defining a second measure, which we can later compare with $s^{NC}_i(\delta_i)$.
First we define the \emph{interlayer-degree} $\tilde{k}_n(\delta_i)$ of an author $n$ as the total number of distinct papers written by $n$ \emph{before} time $\delta_i$.
This measure allows us to quantify the experience of author $n$ which she gained before a given point in time.
To characterize a paper $i$ at time $\delta_i$ by using this information of its authors~$r$ before time $\delta_i$, we compute

\begin{align}
	\label{eq:2}
	s^{NP}_i(\delta_i) = \sum_r \tilde{k}_r(\delta_i) - \tilde{C}(\delta_i)
\end{align}
in analogy to Eq.~\eqref{eq:1}.
Here $\tilde{C}(\delta_i)$ is again a correction term used to only count unique papers (if some authors had written a paper together in the past already).
Its derivation can again be found in Appendix~\ref{sec:correction-terms}.

\paragraph{Parametrising citation rates.}
The quantities $s^{NC}_i(\delta_i)$ and $s^{NP}_i(\delta_i)$ are based on the information of the \emph{authors} of paper $i$.
Our goal is to determine how they influence the \emph{citation dynamics} of paper $i$.
i.e. we need an analytically tractable parametrisation of the citation rates.
To parametrise the citation dynamics we resort to the two characteristic phases of citation histories mentioned in Sect.~\ref{sec:attention:cit-rate}.
The first phase corresponds to increasing citation rates, and we parametrise by its duration $t_i^\text{peak}$, because we have no more precise knowledge about a general functional form of this phase.
The second phase corresponds to an exponential decay, and we parametrise it as the parameter $\tau_i$ in Eq.~\eqref{eq:attention:exp-decay}, i.e. the so-called lifetime.
Both parameters, $t_i^\text{peak}$ and $\tau_i$, are illustrated in Figure~\ref{fig:attention:history-two-phases}.

We now have four parameters to summarize the information about paper $i$.
The first two parameters are $s^{NC}_i(\delta_i)$ and $s^{NP}_i(\delta_i)$, which characterise the \emph{authors} of paper $i$.
The other two parameters are $t_i^\text{peak}$ and $\tau_i$ which characterise the \emph{citation history}.

\paragraph{Excluding incomplete citation histories.}
Obviously our data sets only contain papers published before the release date of the respective database.
Hence, the time-span on which we can compute a given paper's citation history, is also limited by this date.
This introduces an issue especially for recent papers: the observable period of the citation history can be so short that the decay phase has not yet started at all.
To account for this, we omitted all papers that were published within the last 5 years before the release of the respective database.
Hence, for all papers in our study the citation histories are covered over at least 5 years.
In addition, we also removed those papers whose citation rate is non-decreasing in the latest year, since this is a sign that the respective paper has not yet reached its decay phase.

\subsection{Time to the Peak Citation Rate}\label{sec:attention:peak-time}

\paragraph{Regressions.}
Let us first analyse the relationship between the time \(t^{\text{peak}}_i\) until paper \(i\) reaches its highest citation rate, and the social relations of its authors.
In order to find whether there is a significant relationship, we perform a linear analysis for log-transformed variables:

\begin{equation}
    \log_{10} t^{\text{peak}}_i = \beta_0^{\text{peak}} + \beta_1^{\text{peak}} \cdot \log_{10} s_i
    \label{eq:attention:regr:peak}
\end{equation}

where \(s_i\) is the number of previous coauthors $s^{NC}_i$ or the number of previous publications $s^{NP}_i$, and $t^{\text{peak}}_i$ is measured in years.

\paragraph{Results.}
They are presented in Table~\ref{tab:regression-fits}.
All fitted parameters $\beta_1^{\text{peak}}$ are negative and significantly different from $0$ (on a significance level of $0.05$).
But how large is their effect for the untransformed variables?
By exponentiating Eq.~\eqref{eq:attention:regr:peak} we obtain the relation

\begin{equation}\label{eq:time-to-peak-vs-si}
	t^{\text{peak}}_i \sim {[s^{NC}_i]}^{\beta_1^{\text{peak}}}
\end{equation}

In particular, we see that \(\beta^{\text{peak}}_1\) becomes an exponent for the untransformed variables.
Because it is negative, it follows that the time to reach the peak citation rate $t^{\text{peak}}_i$ is \emph{shorter} for papers with \emph{more previous coauthors} (i.e. larger $s^{NC}_i$).
For example, in the case of \gls{jpra} the fitted \(\beta_1^{\text{peak}} = -0.061\) predicts that the time it takes to reach the peak citation rate for a paper whose authors have 100 coauthors, will be on average 34\% faster than for a paper whose authors only have 1 coauthor.
We find the same statistically significant negative effect also for the number of previous publications \(s^{NP}_i\).
This means that the time it takes to reach the peak citation rate also tends to be \emph{shorter} if the authors of the given paper have written \emph{more publications} prior to it.

\begin{table}[htbp]
	\caption{
		\small
		Fitted citation rate parameters \(\beta_1^{\text{peak}}\) and \(\beta_1^{\tau}\) for the examined journals in our APS data set (left) and in our INSPIREHEP data set (right).
		The parameters are the respective coefficients in the linear regressions in Eqs.~\eqref{eq:attention:regr:peak} and \eqref{eq:attention:regr:decay}.
		The significance levels of the p-values for the estimated parameters are encoded as ${}^{***}$ ($<0.001$), ${}^{**}$ ($<0.01$), ${}^*$ ($<0.05$).  }\label{tab:regression-fits}

	\scriptsize
	\centering
	\begin{tabular}[t]{rlll} \toprule
		$s_i$  &  Time  &  \qquad $\beta_1^{\text{peak}}$  &  \qquad $\beta_1^{\tau}$  \\[1ex]
		\midrule
		\multicolumn{4}{@{}l}{\textbf{\gls{jpr}}} \\
		\midrule
		\multirow{2}{*}{NC}
			& years    & $-0.079 \pm 0.009$ ***    & $-0.082 \pm 0.008$ *** \\
			& pubs    & $-0.054 \pm 0.009$ ***    & $-0.013 \pm 0.008$ ~~~ \\
		\multirow{2}{*}{NP}
			& years    & $-0.028 \pm 0.009$ \,**\,    & $-0.032 \pm 0.008$ *** \\
			& pubs    & $-0.035 \pm 0.009$ ***    & $-0.020 \pm 0.008$ ~*~ \\
		\multicolumn{4}{@{}l}{\textbf{\gls{jpra}}} \\
		\midrule
		\multirow{2}{*}{NC}
			& years    & $-0.074 \pm 0.006$ ***    & $-0.138 \pm 0.005$ *** \\
			& pubs    & $-0.014 \pm 0.006$ ~*~    & $-0.056 \pm 0.005$ *** \\
		\multirow{2}{*}{NP}
			& years    & $-0.084 \pm 0.006$ ***    & $-0.142 \pm 0.006$ *** \\
			& pubs    & $-0.019 \pm 0.006$ \,**\,    & $-0.058 \pm 0.005$ *** \\
		\multicolumn{4}{@{}l}{\textbf{\gls{jprc}}} \\
		\midrule
		\multirow{2}{*}{NC}
			& years    & $-0.028 \pm 0.007$ ***    & $-0.071 \pm 0.007$ *** \\
			& pubs    & $-0.010 \pm 0.007$ ~~~    & $-0.052 \pm 0.007$ *** \\
		\multirow{2}{*}{NP}
			& years    & $-0.054 \pm 0.009$ ***    & $-0.083 \pm 0.009$ *** \\
			& pubs    & $-0.041 \pm 0.009$ ***    & $-0.063 \pm 0.009$ *** \\
		\multicolumn{4}{@{}l}{\textbf{\gls{jpre}}} \\
		\midrule
		\multirow{2}{*}{NC}
			& years    & $-0.029 \pm 0.007$ ***    & $-0.041 \pm 0.007$ *** \\
			& pubs    & $-0.021 \pm 0.008$ \,**\,    & $-0.049 \pm 0.008$ *** \\
		\multirow{2}{*}{NP}
			& years    & $-0.045 \pm 0.008$ ***    & $-0.036 \pm 0.008$ *** \\
			& pubs    & $-0.035 \pm 0.009$ ***    & $-0.038 \pm 0.009$ *** \\
		\multicolumn{4}{@{}l}{\textbf{\gls{jrmp}}} \\
		\midrule
		\multirow{2}{*}{NC}
			& years    & $-0.167 \pm 0.057$ \,**\,    & $-0.272 \pm 0.054$ *** \\
			& pubs    & $-0.155 \pm 0.049$ \,**\,    & $-0.337 \pm 0.051$ *** \\
		\multirow{2}{*}{NP}
			& years    & $-0.178 \pm 0.071$ ~*~    & $-0.247 \pm 0.070$ *** \\
			& pubs    & $-0.169 \pm 0.057$ \,**\,    & $-0.281 \pm 0.062$ *** \\
		\bottomrule
	\end{tabular}
	\qquad
	\qquad
	\begin{tabular}[t]{rlll} \toprule
		$s_i$  &  Time  &  \qquad $\beta_1^{\text{peak}}$  &  \qquad $\beta_1^{\tau}$  \\[1ex]
		\midrule
		\multicolumn{4}{@{}l}{\textbf{\gls{jhep}}} \\
		\midrule
		\multirow{2}{*}{NC}
			& years    & $-0.050 \pm 0.014$ ***    & $-0.061 \pm 0.009$ *** \\
			& pubs    & $-0.050 \pm 0.013$ ***    & $-0.002 \pm 0.008$ ~~~ \\
		\multirow{2}{*}{NP}
			& years    & $-0.023 \pm 0.012$ ~~~    & $-0.026 \pm 0.008$ *** \\
			& pubs    & $-0.033 \pm 0.011$ \,**\,    & $\hphantom{-}0.011 \pm 0.007$ ~~~ \\
		\multicolumn{4}{@{}l}{\textbf{\gls{jpr_ins}}} \\
		\midrule
		\multirow{2}{*}{NC}
			& years    & $-0.129 \pm 0.008$ ***    & $-0.159 \pm 0.007$ *** \\
			& pubs    & $\hphantom{-}0.011 \pm 0.009$ ~~~    & $-0.054 \pm 0.008$ *** \\
		\multirow{2}{*}{NP}
			& years    & $-0.125 \pm 0.008$ ***    & $-0.135 \pm 0.007$ *** \\
			& pubs    & $-0.008 \pm 0.009$ ~~~    & $-0.039 \pm 0.007$ *** \\
		\multicolumn{4}{@{}l}{\textbf{\gls{jpl}}} \\
		\midrule
		\multirow{2}{*}{NC}
			& years    & $-0.068 \pm 0.018$ ***    & $-0.080 \pm 0.016$ *** \\
			& pubs    & $-0.101 \pm 0.018$ ***    & $-0.123 \pm 0.016$ *** \\
		\multirow{2}{*}{NP}
			& years    & $-0.074 \pm 0.016$ ***    & $-0.070 \pm 0.014$ *** \\
			& pubs    & $-0.092 \pm 0.016$ ***    & $-0.102 \pm 0.014$ *** \\
		\multicolumn{4}{@{}l}{\textbf{\gls{jnp}}} \\
		\midrule
		\multirow{2}{*}{NC}
			& years    & $-0.085 \pm 0.011$ ***    & $-0.117 \pm 0.010$ *** \\
			& pubs    & $-0.011 \pm 0.011$ ~~~    & $-0.116 \pm 0.010$ *** \\
		\multirow{2}{*}{NP}
			& years    & $-0.094 \pm 0.010$ ***    & $-0.119 \pm 0.008$ *** \\
			& pubs    & $-0.004 \pm 0.010$ ~~~    & $-0.111 \pm 0.009$ *** \\
		\bottomrule
	\end{tabular}
\end{table}

\subsection{Characteristic Decay Time}\label{sec:attention:decay}

\paragraph{Regressions.}
To investigate the characteristic decay time $\tau_i$, we again perform a linear analysis for log-transformed variables.
This corresponds to the following model:

\begin{equation}
    \log_{10} \tau_i = \beta_0^{\tau} + \beta_1^{\tau} \cdot \log_{10} s_i
    \label{eq:attention:regr:decay}
\end{equation}

where again $s_i$ is the number of previous coauthors $s^{NC}_i$ or the number of previous publications $s^{NP}_i$, and the time-unit is again chosen as years.

\paragraph{Results.}
They are presented in Table~\ref{tab:regression-fits}.
Also here all fitted parameters $\beta_1^{\tau}$ are negative and significantly different from $0$ (on a significance level of $0.05$).
To interpret the effect of $s_i$ on $\tau_i$, one can exponentiate Eq.~\eqref{eq:attention:regr:decay} to obtain:

\begin{equation}
	\tau_i \sim [s_i]^{\beta_1^\tau}
\end{equation}

Because $\beta_1^{\tau}$ is negative, this means that the more previous coauthors the authors have, the smaller the value of $\tau_i$ becomes.  From Eq.~\eqref{eq:attention:exp-decay} we know that the smaller $\tau_i$ is, the faster is the decay of the normalized citation rate $\crate_i(t)$.
This in turn means that such a paper faces a quicker and stronger shortage in new citations.
Again, we also find significantly negative parameters $\beta_1^{\tau}$ when using the number of previous publications \(s^{NP}_i\) in Eq.~\eqref{eq:attention:regr:decay}.

\subsection{Rescaling Time by Counting Publications}

\paragraph{Effect of the growing scientific output.}
It is known that the number of papers published every year grows exponentially over time \citep{Price1951}.
This means that in recent years there are more papers published in a given time-interval than this was the case longer ago.
And all of these new publications can potentially cite a given paper.
This time-dependence likely affects our regression results by confounding the respective response (\(t^{\text{peak}}_i\) or $\tau_i$) and predictor variable (\(s^{NC}_i\) or \(s^{NP}_i\)).
In the past it was suggested that the dependence of the citation rate on the publication year of a paper can be weakened by counting time in terms of the number of published papers instead of absolute time (days, weeks, years, etc.) \citep{Parolo2015}.
Therefore we repeat our regressions from Sects.~\ref{sec:attention:peak-time}, and \ref{sec:attention:decay} while measuring time on this alternative timescale.
Thereby we assess whether such a bias from the publication year of a paper is present in the relations which we found.

\paragraph{Results for the alternative timescale.}
These fitted parameters are also listed in Table~\ref{tab:regression-fits}.
For all but two journals they remain smaller than $0$, just like before.
The exceptions are \gls{jhep} whose parameter $\beta_1^\tau$ becomes positive when measured for \(s^{NP}_i\) as independent variable, and \gls{jpr_ins} whose parameter \(\beta_1^\text{peak}\) becomes positive for \(s^{NP}_i\).
However, neither of these two parameters is significantly different from $0$ anymore on a significance level of $0.05$, and hence they do not contradict our previous findings.
Overall, the parameter \(\beta_1^{\text{peak}}\) is significant in all journals except \gls{jpr_ins}, \gls{jnp}, and \gls{jprc}, and the parameter \(\beta_1^{\tau}\) is significant in all journals except \gls{jhep} and \gls{jpr}.
Hence we conclude that time does not introduce a general bias to our findings.

This means that, also according to the alternative timescale, the peak in the citation rate is reached faster for papers by authors with more previous coauthors or publications.
Accordingly, the decay becomes steeper for papers by such authors.
Furthermore, we found this relation to be significant in most journals in our sample, and hence to be a general pattern.  However, statistical significance is of limited interest when studying such large data sets.
Therefore, as a next step, we measure also the size of this effect.

\subsection{Effect Size}\label{sec:effect-size}

\paragraph{Size of peak delays.}
To study the size of the dependence between peak delays, $t_i^\text{peak}$, and the number of previous coauthors, $s^{NC}_i$, or publications, $s^{NP}_i$, we can not use our linear models from Eq.~\eqref{eq:attention:regr:peak}.
This is because we found that the error terms of these regressions are not normally distributed, which they need to fulfil the model assumptions of linear regression, see Appendix~\ref{sec:linear-regression-validation}.
This does not pose a problem for the study of the \emph{significance} of parameters as in our previous sections \citep{Frost2014}, but to assess the \emph{size} of the dependence, these models are no longer applicable.
The reason why the errors are not normally distributed, is that peak-delays are essentially counts.
We count whether the peak citation rate occurs in year 0, or in year 1, or in year 2, etc. after publication.
To predict such count data, linear regression is not suitable, mostly because it predicts the dependent variable on a continuous scale.
Furthermore, because our slope parameters $\beta_1^{\text{peak}}$ are negative, linear regression will predict negative peak delays for large numbers of previous coauthors or publications.

To get around these limitations, we instead apply a negative binomial regression, which is a standard model for count data \citep{Hilbe2011}.
The model we fit is:

\begin{equation}\label{eq:effect-size-peak-regression-equation}
	\begin{aligned}
		t_i^\text{peak} &\sim \text{Poisson}(\lambda_i)\\
		\lambda_i &= f_i{(\alpha + \beta\cdot s_i)}
	\end{aligned}
\end{equation}

As in Sect.~\ref{sec:attention:peak-time}, $s_i$ is the number of previous coauthors, $s_i^{NC}$, or the number of previous publications, $s_i^{NP}$, and the parameters to be fitted are $\alpha$ and $\beta$.
The function $f_i$ is an exponential function containing a random term that accounts for the observed large variance in our data, see Appendix~\ref{sec:binomial-regression-validation}.
We leave its derivation to dedicated literature such as \citep{Hilbe2011}.
The applicability of this negative binomial regression for our data is validated in Appendix~\ref{sec:binomial-regression-validation}.
To fit Eq.~\ref{eq:effect-size-peak-regression-equation} we used the function \emph{glm.nb} in the R-package \emph{MASS}.
The important difference from the linear regression model, Eq.~\ref{eq:attention:regr:peak}, is that the negative binomial regression model, Eq.~\ref{eq:effect-size-peak-regression-equation}, uses a discrete (Poisson) distribution for $t_i^\text{peak}$.

In Table~\ref{tab:effect-size-peak-regression-table} we show the fitted coefficients for all journals.
Except for one, all coefficients $\beta$ are negative, which confirms our previous finding of decreasing peak delays for increasing numbers of previous coauthors or publications.
The exception is \gls{jhep} which has a positive $\beta$ for $s_i$ as the number of previous publications, $s_i^\text{NP}$.
However, this coefficient is not significant, meaning that it is likely not different from zero, and therefore does not contradict the discovered trend.

\begin{table}
	\caption{
		Fitted coefficients $\alpha$, $\beta$ for the negative binomial regression in Eq.~\eqref{eq:effect-size-peak-regression-equation}, computed for each journal individually.
		\textit{NC} means that the number of previous coauthors $s_i^{NC}$ is used as the predictor $s_i$, and \textit{NP} accordingly means the number of previous publications $s_i^{NP}$.
		The stated significance levels concern the estimated parameters $\beta$, and are given as ${}^{***}$ ($<0.001$), ${}^{**}$ ($<0.01$), ${}^*$ ($<0.05$).  }
	\label{tab:effect-size-peak-regression-table}

	\small
	\centering
    \begin{tabular}{l|ccc|ccc|}
        &  \multicolumn{3}{c|}{\textit{NC}} & \multicolumn{3}{c|}{\textit{NP}} \\
                             &  $\alpha$  &  $\beta$   &  Sign.  &  $\alpha$  &  $\beta$   &  Sign.   \\
        \midrule
		\acrshort{j0}        &  $\hphantom{-}0.768$   &  $-0.022$  &  ***    &  $\hphantom{-}0.721$  &  $-0.010$  &  ***     \\
		\acrshort{j1}        &  $\hphantom{-}1.055$   &  $-0.001$  &  ***    &  $\hphantom{-}1.095$  &  $-0.005$  &  ***     \\
        \acrshort{j4}        &  $\hphantom{-}1.141$   &  $-0.000$  &  *      &  $\hphantom{-}1.130$  &  $-0.000$  &  ~~~     \\
		\acrshort{j6}        &  $\hphantom{-}0.907$   &  $-0.001$  &  ***    &  $\hphantom{-}0.941$  &  $-0.004$  &  ***     \\\acrshort{j12}       &  $\hphantom{-}1.987$   &  $-0.004$  &  *      &  $\hphantom{-}2.022$  &  $-0.008$  &  *       \\\midrule
        \acrshort{j1213103}  &  $\hphantom{-}0.050$   &  $-0.002$  &  **     &  $-0.035$  &  $\hphantom{-}0.000$  &  ~~~     \\
        \acrshort{j1214516}  &  $\hphantom{-}1.292$   &  $-0.005$  &  ***    &  $\hphantom{-}1.321$  &  $-0.004$  &  ***     \\
        \acrshort{j1214521}  &  $\hphantom{-}0.813$   &  $-0.007$  &  ***    &  $\hphantom{-}0.824$  &  $-0.004$  &  ***     \\
        \acrshort{j1214548}  &  $\hphantom{-}1.156$   &  $-0.007$  &  ***    &  $\hphantom{-}1.211$  &  $-0.005$  &  ***     \\
        \bottomrule
    \end{tabular}
\end{table}

\begin{figure}
	\centering
	{\tiny\effectslegend{}} \newline
	\includegraphics[width=0.46\textwidth]{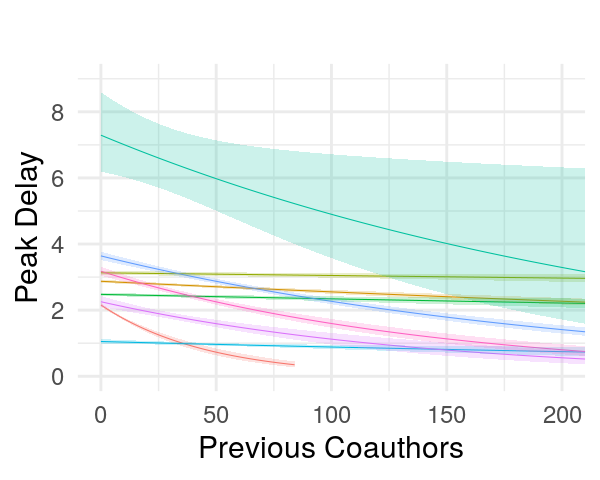}
	\hfill
	\includegraphics[width=0.46\textwidth]{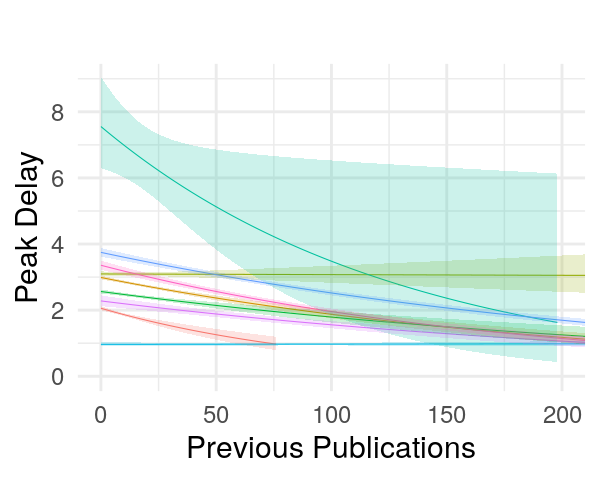}
	\caption{
		Relation between peak delays $t_i^\text{peak}$ measured in years, and the number of previous coauthors $s_i^{NC}$ (left) or publications $s_i^{NP}$ (right) according to Eq.~\eqref{eq:effect-size-peak-regression-equation}.
		The solid lines are the estimated responses, and the respective colored areas are $95\%$ confidence bands from the negative binomial regressions.
		Estimated responses are plotted at most until the largest observed number of previous coauthors or publications in the respective journal.
	}
	\label{fig:effect-size-plots-peak}
\end{figure}

Our main aim behind the fitted models listed in Table~\ref{tab:effect-size-peak-regression-table} is to study the size of the peak delay for given numbers of previous coauthors or publications.
Specifically, we use these models to predict average peak-delays for given numbers of previous coauthors, $s_i^{NC}$, or publications, $s_i^{NP}$.
In Figure~\ref{fig:effect-size-plots-peak} we visualize these predictions for all journals.  Let us first focus on the number of previous coauthors $s_i^{NC}$, Figure~\ref{fig:effect-size-plots-peak} (left).
We see that for all journals except \acrshort{j12} the predicted average $t_i^\text{peak}$ is always smaller than 4 years irrespective of the number of previous coauthors.
For \acrshort{j12} papers with 0 authors take on average around 7.5 years to reach the peak, but this number then also decreases to 4 years at roughly 150 previous coauthors.

We further point out the differences in speed across journals, at which the peak delays decrease for increasing numbers of previous coauthors.
For example, papers in the journal \acrshort{j1214516} reach the peak citation rate on average after 3.75 years for 0 previous coauthors, while this changes to roughly 2.5 years for papers with 100 previous coauthors.
This is different from the journal \acrshort{j6} where a paper reaches the peak citation rate on average after 2.5 years for 0 previous coauthors, while this number stays almost the same even at 100 previous coauthors.
This means that journals have a large impact on the time \emph{when} citations occur, especially with respect to the prospective decrease as the number of coauthors grows.
Figure~\ref{fig:effect-size-plots-peak} also shows confidence bands for the predicted average $t_i^\text{peak}$.
These are narrow for all journals except \acrshort{j12}, because of the large numbers of papers used in the model fits.
For \acrshort{j12} only 214 papers were used, and hence its confidence bands are wider.

Figure~\ref{fig:effect-size-plots-peak} (right) shows the average $t_i^\text{peak}$ predicted by the number of previous publications, $s_i^{NP}$.
The main difference to Figure~\ref{fig:effect-size-plots-peak} (left) is that now also the journals \acrshort{j1} and \acrshort{j6} decrease noticeably for increasing numbers of previous publications.
More precisely, $t_i^\text{peak}$ is on average roughly 3 years for 0 previous publications, while this number now drops to 1 year for 200 previous publications.
This means that, to receive citations earlier in these journals, increasing the number of publications appears to be a more successful strategy than to increase the number of coauthors.

To summarize, the negative binomial regression models confirm that for \emph{increasing} numbers of previous coauthors or publications the highest citation rate is reached \emph{sooner}.
They also identify differences in the benefit of high numbers of coauthors or publications across journals: For journals like \acrshort{j4} there is almost no decrease in peak delay even with 200 previous coauthors.
But for journals like \acrshort{j0}, papers with already 50 previous coauthors reach their peak on average in less than half the time of papers with 0 previous coauthors.

\begin{figure}
	\centering
	{\tiny\effectslegend{}} \newline
	\includegraphics[width=0.46\textwidth]{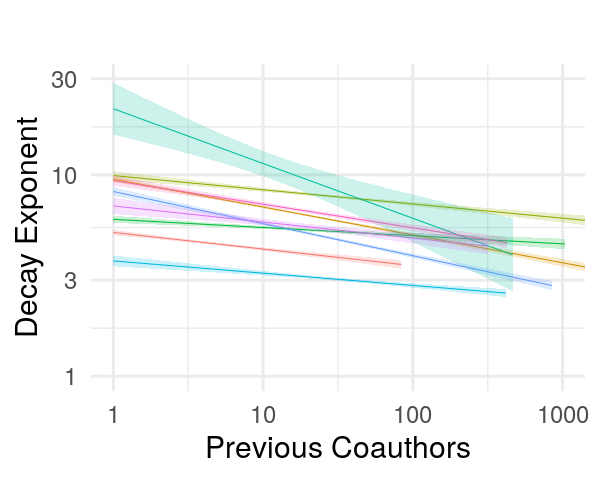}
	\hfill
	\includegraphics[width=0.46\textwidth]{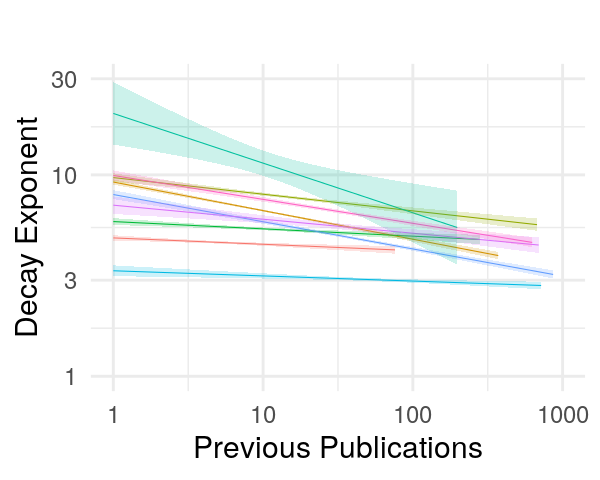}
	\caption{
		Relation between decay exponents $\tau_i$ and the number of previous coauthors $s_i^{NC}$ (left) or the number of previous publications $s_i^{NP}$ (right) according to Eq.~\eqref{eq:attention:regr:decay}.
		The solid lines are the estimated responses, and the respective colored areas are $95\%$ confidence bands derived from the standard errors.
		Estimated responses are plotted at most until the largest observed number of previous coauthors or publications in the respective journal.
	}
	\label{fig:effect-size-plots-decay}
\end{figure}

\paragraph{Size of decay exponents.}
To study the size of the dependence between decay exponents $\tau_i$ and the number of previous coauthors $s^{NC}_i$ or publications $s^{NP}_i$, we apply our regression models from Sect.~\ref{sec:attention:decay}.
In Appendix~\ref{sec:binomial-regression-validation} we show that the assumptions of linear regression models are reasonably fulfilled to allow for predictions.
In Figure~\ref{fig:effect-size-plots-decay} we visualize the estimated average decay parameters for the different journals.
We focus on the description of the number of previous coauthors, Figure~\ref{fig:effect-size-plots-decay} (left), because overall both plots convey a similar message.
We see that for papers with 0 previous coauthors the decay exponents are below 10 for all journals, except for \acrshort{j12} which attains a decay exponent below 30.
We further point out that papers in the journal \acrshort{j1213103} have the smallest decay exponents for up to even 1000 previous coauthors.
This in turn means that decays in this journal tend to be particularly fast compared to the other journals.

\section{Conclusions}\label{sec:attention:conclusion}

In this paper, we address the question how the attention towards an academic publication is accumulated over time, depending on the social relations of its authors, as expressed in the coauthorship network.
For example, does the attention mostly occur in an early phase right after publication?
Or is it rather spread uniformly over time?
Or might it even happen only after a long time has passed since publication?
To obtain a tractable, objective characterisation of attention, we proxy attention by the citation rate of a paper, i.e. the number of new citations obtained in a particular time-interval.
We argue that, in order for a citation to occur, the authors of the citing paper have to be aware of the cited paper.

To study the time when this attention occurs, we compute the change in the number of citations over a time-interval, i.e. the citation rate.
It is known that citation rates of most papers have two characteristic phases over time, namely an increasing phase followed by a decay phase.
We found that the first phase tends to get shorter and the decay in the second phase tends to get faster for papers written by authors who have many previous coauthors.
We also found that for some journals the time to the peak citation rate is almost halved within the first 100 previous coauthors, while for other journals it stays almost unchanged.
Such a difference is also present in the decay exponents for different journals.

In terms of attention, our findings mean that papers written by authors with more previous coauthors attract attention faster, but are then also forgotten sooner.
We also found this effect when measuring the number of previous papers of the authors instead of the number of previous coauthors.
Furthermore, this effect also persisted when we controlled for the time when a paper was published.  But most importantly, we found this effect in \emph{9 journals}, based on \emph{hundreds of thousands of authors and papers} and \emph{far more than a million of citations}.
A study of such a large scale is a strong sign that we uncovered a general trend which is not limited to the analysed data sets.

\paragraph{A speculative explanation.}
Which mechanisms could be responsible for this?
One way how authors learn about the papers which they cite is through communication with other scientists.
Hence, authors can use their (few or many) social contacts, proxied by coauthors, to ``advertise'' a paper.
Our findings indicate that authors with many previous coauthors or papers tend to do so within a short period of time after publication.
When a new publication is made, the authors ``advertise'' it to the scientific community by presenting it in conferences and seminars, by sharing it on social media, etc.
This behaviour happens within a finite time period, after which the authors stop actively promoting the given publication.
However, this explanation is merely speculative at this point.

\paragraph{Regressions not suitable for predictions.}
Our performed regressions have low predictive power, as indicated by extremely small coefficients of determination, $R^2$.
For instance, for some regressions the $R^2$ is as low as $0.001$, meaning that only 0.1\% of the variance in the dependent variable is explained.
However, while our regression models are not useful for prediction, our inferred relations are significant.
In particular our regressions show that the time to the peak citation rate and the subsequent decay are not independent of the authors.

\paragraph{Future work.}
Our analysis shows that the social relations of authors significantly influence the citation histories of their papers.
But our analysis does not provide insights about the \emph{mechanisms} behind this influence.
In the future, we can use generative modelling to learn more about these underlying mechanisms.
For instance, hypotheses can be formulated and tested using the framework of coupled growth models presented in \citep{Nanumyan2020}.

We find that a paper receives attention from the scientific community faster, the more coauthors the authors had prior to its publication.
But we find as well that such a paper is also forgotten sooner again afterwards.
Our findings indeed highlight that the citations of a paper can have substantially different dynamics depending on the social relations of the authors.
Furthermore, our approach illustrates how such coupled dynamics can be studied by representing scientific collaborations in a multilayer network.

\section*{Acknowledgements}
The authors acknowledge discussions with Luca Verginer and Giacomo Vaccario concerning the negative binomial regression models.

\bibliographystyle{apa}
\bibliography{bibliography}

\begin{appendix}
	
	\section{Correction Terms}\label{sec:correction-terms}

	$s^{NC}_i(\delta_i)$ is defined in Eq.~\eqref{eq:1} as the number of persons on the coauthorship layer who had co-authored at least one publication with any of the authors of paper $i$ before time~$\delta_i$.
	The aim of the correction-term $C(\delta_i)$ in Eq.~\eqref{eq:1} is to correct the sum of the authors' degrees by those co-authors who collaborated with more than just one author.
	I.e., so that $s_i^{NC}(\delta_i)$ does not count some co-authors multiple times.
	Precisely, $C(\delta_i)$ can be computed as

	\begin{equation}
		\label{eq:1Correction}
		C(\delta_i) = \max\left\{0, \sum_{s\in V^a} \left(\sum_{r} 1_{s,r}(\delta_i) - 1\right)\right\}
	\end{equation}

	where $s$ is any person on the coauthorship layer, $r$ is an author of paper $i$, and $1_{s,r}(\delta_i)$ is equal to $1$ exactly if $s$ and $r$ have coauthored at least one paper before time $\delta_i$, and is $0$ otherwise.
	The intuition of Eq.~\eqref{eq:1Correction} is as follows.
	The sum over $r$ counts for a specific person $s$ in the coauthorship layer with how many of the authors of paper $i$ she or he has at least one previous coauthorship before time $\delta_i$.
	This sum is exactly equal to the number of times the sum of the degrees in Eq.~\eqref{eq:1} counts person $r$ in total.
	But now we have to subtract $1$ from this number, because otherwise we are removing person $s$ completely from the sum of the degrees, whereas she or he should be contained with a weight of exactly $1$.
	The ``$\max\{0, \dots\}$'' is a numerical trick to handle those persons~$s$ correctly who did not collaborate with any of the authors of paper $i$ before time $\delta_i$.
	If we would omit this part, they would add a negative weight to the correction term, instead of $0$.
	Thereby, Eq.~\eqref{eq:1Correction} yields the correction necessary to adjust Eq.~\eqref{eq:1} into the number of persons on the co-authorship layer who have collaborated with at least one author before time $\delta_i$.

	In analogy to Eq.~\eqref{eq:1Correction} the correction term for Eq.~\eqref{eq:2} can be computed as

	\begin{equation}
		\label{eq:2Correction}
		\tilde{C}(\delta_i) = \max\left\{0, \sum_{p\in V^p}\left(\sum_r 1_{p,r}(\tau_i) - 1\right)\right\}
	\end{equation}

	where $p$ is any paper on the citation layer published before time $\delta_i$, $r$ is an author of paper $i$, and $1_{p,r}(\delta_i)$ is equal to $1$ exactly if $r$ is an author of paper $p$.

	\section{Linear Regression Model Validation}\label{sec:linear-regression-validation}

	\subsection{Peak Delay Models}

	\paragraph{Aspects to validate.}  For a linear regression to be valid, the residuals must be normally distributed, their variance must not depend on the explanatory variables, and their expectation must be zero.  To confirm that our regressions according to Eq.~\eqref{eq:attention:regr:peak} are valid, we look at established diagnostic tools for these assumptions.
	We show a validation here for the example of the journal \acrshort{j1}, $s_i$ as the number of previous coauthors $s_i^{NC}$, and time measured in years.

\begin{figure}[htbp]
		\centering
		\includegraphics[width=0.32\textwidth]{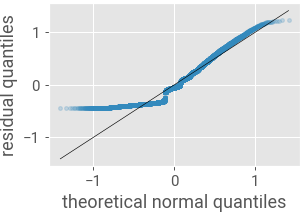}
		\includegraphics[width=0.32\textwidth]{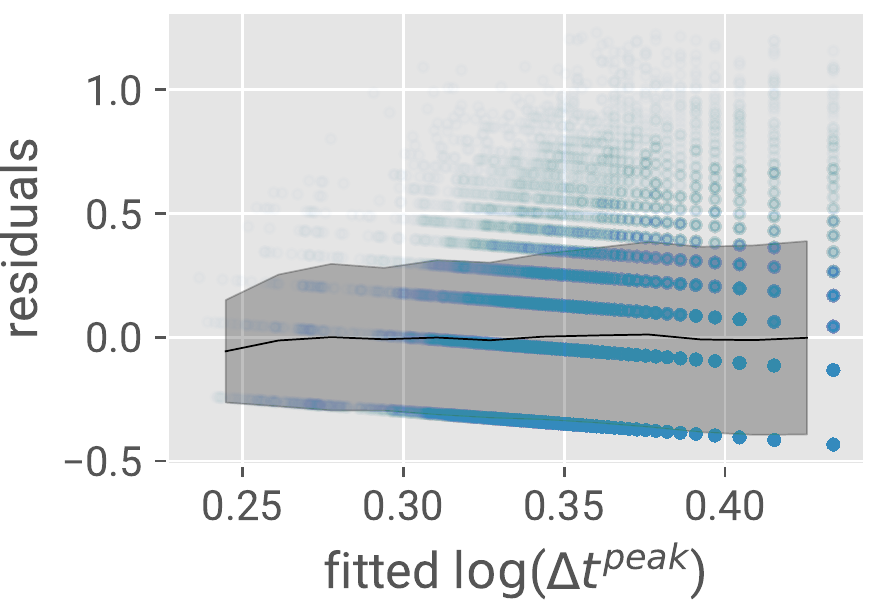}
		\includegraphics[width=0.32\textwidth]{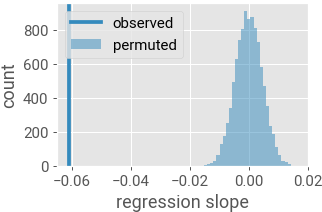}
		\caption{Validity of Eq.~\eqref{eq:attention:regr:peak} for \gls{jpra}. (Left) quantile-quantile plot of residuals versus normal distribution, (middle) Tukey-Anscombe plot and (right) the result of permutation test based on resampling the data 10000 times.
			}\label{fig:attention:peak-diag}
	\end{figure}

	\paragraph{QQ-plot.}
	To test for normality, we look at the Quantile-Quantile (QQ) plot between the observed distribution of the residuals and a theoretical normal distribution.
	If the observed distribution is the same as the theoretical one, the points in the QQ-plot will all fall close to the identity line.
	This plot is shown for the example in the left panel of Figure~\ref{fig:attention:peak-diag}.
	We see that in the lower tail, i.e. for the smallest negative residuals, the lowest observed quantile stretches over almost the whole negative range of the theoretical quantiles.
	Hence, the plot indicates a violation of the normality assumption in our case.
	This violation is due to the finite size of the time unit over which we have computed the citation rates, i.e., the peak delays are essentially counts and not a continuous variable.
	However, non-normality of the residuals does not threaten the significance of the slope~\citep{Frost2014}, which is the use-case in our trend detection in Section~\ref{sec:attention:peak-time}.
	Yet still, it means that these regressions misestimate the size of the involved variables, and that therefore we need a different model for our study of effect size in Section~\ref{sec:effect-size}.
	For this task we will instead use negative binomial regressions.

	\paragraph{Tukey-Anscombe plot.}
	This plot shows the residuals against the predicted values of the dependent variable, and is used to check whether the expectation of the residuals is zero and independent from the explanatory variable.
	In the middle panel of Figure~\ref{fig:attention:peak-diag} we present the Tukey-Anscombe plot for our example case.
	The black line is the mean of the residuals for different values of the dependent variable.
	We see that it is close to zero for all values of the predicted dependent variable.

	Based on the Tukey-Anscombe plot, we can also check the condition of homoskedasticity, i.e., that the variances of the residuals are constant.
	To this end, we include the standard error of the residuals against the predicted value of the dependent variable.
	We see that for small values of fitted peak-delays it grows slightly with the dependent variable.
	However, the estimation of standard errors in that region is not very reliable due to the relatively few observed data points.
	We conclude that overall the assumption of constant variance is well satisfied.

	\paragraph{Permutation test.}
	The mentioned plots allow only a qualitative evaluation of the validity of the regressions.
	As we are interested in the statistical significance of the dependence between the dependent and explanatory variables, we confirm the significance of the slope by means of a permutation test.
	That is, we randomly reshuffle the values of the explanatory variable between different papers, while keeping the original values of the dependent variable.
	Then we perform the regression analysis on this shuffled data.
	By repeating this procedure multiple times, we obtain a distribution of values of the regression analysis based on the randomised data.
	The right panel of Figure~\ref{fig:attention:peak-diag} shows the outcome for the slope of the regression for $10000$ randomised trials.
	We find that the distribution of the slope \(\beta_1^{\text{peak}}\) is centred around zero, meaning that there is no randomly expected effect merely from how the explanatory or dependent variables are distributed.
	We also find that the estimated slope from the observed data is far outside the distribution of the randomised data.
	This confirms the statistical significance of the identified dependence between the time to the peak citation rate and the number of previous coauthors of the authors.
	Hence, we can conclude that the social relations of authors have a significant influence on the time $t^text{peak}_i$ it takes a paper $i$ to reach the largest citation rate.

	\subsection{Decay Models}
	
	Let us now check the validity of the linear regression analysis for the regression in Eq.~\eqref{eq:attention:regr:decay}.
	We again show here the example of the journal \acrshort{j1}, $s_i$ as the number of previous coauthors $s_i^{NC}$, and time measured in years.

\begin{figure}[htbp]
		\centering
		\includegraphics[width=0.32\textwidth]{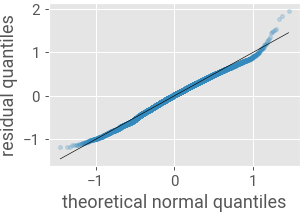}
		\includegraphics[width=0.32\textwidth]{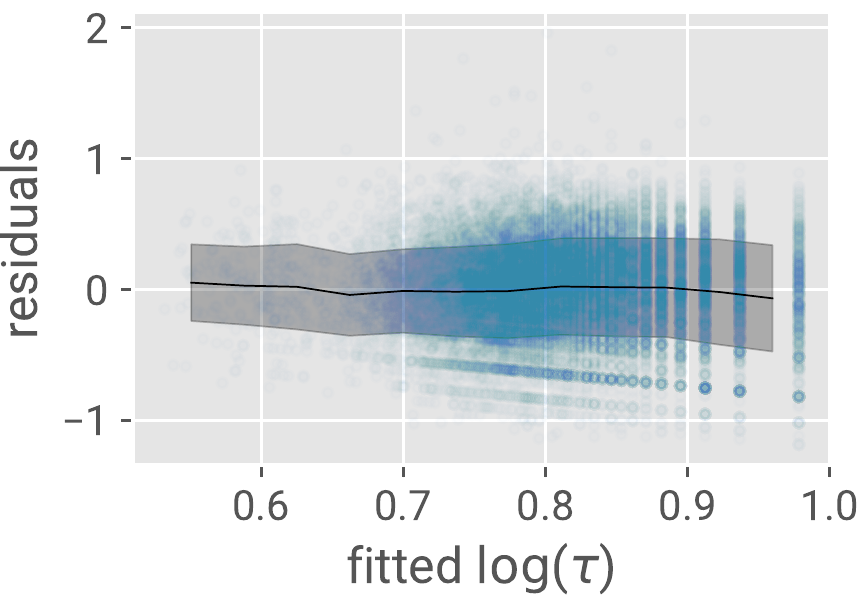}
		\includegraphics[width=0.32\textwidth]{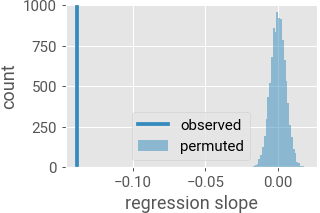}
		\caption{Validity of Eq.~\eqref{eq:attention:regr:decay} for \gls{jpra}. (Left) quantile-quantile plot of residuals versus normal distribution, (middle) Tukey-Anscombe plot and (right) the result of permutation test based on resampling the data 10000 times.
		}\label{fig:attention:decay-diag}
	\end{figure}

	\paragraph{QQ-plot.}
	It is presented in the left panel of Figure~\ref{fig:attention:decay-diag}.
	We see that overall the points coincide well with the diagonal.
	Only in the tails, above 1 standard deviation and below -1 standard deviation, the residual quantiles tend to be slightly larger than those by the theoretical normal distribution.
	However, in these regions there are also only few observed points.
	Hence, we conclude that overall the vast majority of residuals is adequately described by a normal distribution.

	\paragraph{Tukey-Anscombe plot.}
	Next, we inspect the Tukey-Anscombe plot, which is shown in the middle panel of Figure~\ref{fig:attention:decay-diag}.
	We find that the means of the residuals are very close to zero for all values of the predicted decay exponents.
	Furthermore, also the standard deviation is almost constant for different fitted decay exponents.
	We conclude that all assumptions of linear regression are reasonably met for the regression in Eq.~\eqref{eq:attention:regr:decay}.

	\paragraph{Permutation test.}
	To confirm the statistical significance that there is a dependence between the dependent and explanatory variables, we again also perform a permutation test by randomly reshuffling the values of the explanatory variable \(s^{NC}_i\) between papers.
	Performing the regression on 10000 such randomisations, we obtain the distribution of the slope \(\beta_1^{\tau}\) shown in the right panel of Figure~\ref{fig:attention:decay-diag}.
	We see that the slope inferred for the empirical data is far from the distribution from the randomised data.
	Hence, we can conclude that the number of previous coauthors has a significant influence on the characteristic decay time of a paper.
	The more coauthors the authors of a given paper have, the shorter is the characteristic decay time of the paper's citation rate.
	
	\section{Binomial Regression Model Validation}\label{sec:binomial-regression-validation}

	In our approach peak-delays, $t_i^\text{peak}$, are computed on a year-timescale.
	For example, a peak delay can be 0, or 1, or 2, etc., depending on whether the peak is reached in less than one year, or in year 1, or in year 2, etc.
	Thereby, our peak-delays are essentially counts, for which Poisson regression or negative binomial regression are typical models.
	Since it can take many years for papers to get discovered by the scientific community, and hence receive citations, we expect that the variance in peak delays across papers is large.
	If the variance of the peak-delays is larger than their mean, an assumption of Poisson regression is violated, while negative binomial regression becomes applicable.
	To test this assumption, we test the Null Hypothesis

	\begin{equation}\label{eq:overdispersion-h0}
		H_0: Var[t_i^\text{peak} \,| \,s_i] = E[t_i^\text{peak} \,| \,s_i]
	\end{equation}

	against the Alternative Hypothesis

	\begin{equation}\label{eq:overdispersion-ha}
		H_A: Var[t_i^\text{peak} \,| \,s_i] > E[t_i^\text{peak} \,| \,s_i]
	\end{equation}

	To test these Hypotheses we apply the overdispersion test introduced in \citep{Cameron1990}, see their Section~2.
	To compute this test, we applied the function \emph{dispersiontest} from the R-package \emph{AER}.
	The obtained p-values are displayed in Table~\ref{tab:overdispersion-tests}.
	We find that they are all highly significant, implying that $H_0$ should be rejected in favour of $H_A$.
	Hence, we choose negative binomial regression as our model for peak-delays.

	\begin{table}
		\caption{
			The p-values obtained for the overdispersion test from \citep{Cameron1990}, computed for each journal and $s_i$ separately.
			\textit{NC} means that $s_i^\text{NC}$ is used as $s_i$ in Eqs.~\eqref{eq:overdispersion-h0}, \eqref{eq:overdispersion-ha}.
			Accordingly, \textit{NP} means that $s_i^\text{NP}$ is used as $s_i$ in these Eqs.
			\label{tab:overdispersion-tests}
		}
		\centering
		\small
		\hfill
		\begin{tabular}[t]{lcll}
			\toprule
				& $s_i$ & \qquad p-value \\
			\midrule
			\gls{jpr} & NC & $2.85\times 10^{-104}$ & *** \\
				& NP & $5.72\times 10^{-94}$ & *** \\
			\gls{jpra} & NC & $1.80\times 10^{-191}$ & *** \\
				& NP & $5.34\times 10^{-193}$ & *** \\
			\gls{jprc} & NC & $1.07\times 10^{-130}$ & *** \\
				& NP & $4.51\times 10^{-129}$ & *** \\
			\gls{jpre} & NC & $1.51\times 10^{-164}$ & *** \\
				& NP & $9.21\times 10^{-168}$ & *** \\
			\gls{jrmp} & NC & $6.04\times 10^{-8}$ & *** \\
				& NP & $1.37\times 10^{-7}$ & *** \\
			\bottomrule
		\end{tabular}
		\hfill
		\begin{tabular}[t]{lcll}
			\toprule
				& $s_i$ & \qquad p-value \\
			\midrule
			\gls{jhep} & NC & $1.11\times 10^{-83}$ & *** \\
				& NP & $3.18\times 10^{-83}$ & *** \\
			\gls{jpr_ins} & NC & $1.13\times 10^{-113}$ & *** \\
				& NP & $6.68\times 10^{-117}$ & *** \\
			\gls{jpl} & NC & $2.29\times 10^{-30}$ & *** \\
				& NP & $1.14\times 10^{-30}$ & *** \\
			\gls{jnp} & NC & $2.90\times 10^{-94}$ & *** \\
				& NP & $8.45\times 10^{-94}$ & *** \\
			\bottomrule
		\end{tabular}
		\hfill
		\mbox{}
	\end{table}
\end{appendix}

\end{document}